# Super strong paramagnetism of aromatic peptides adsorbed with monovalent cations

Shiqi Sheng[1], Haijun Yang[2,3], Liuhua Mu[3,9], Zixin Wang[5], Jihong Wang[1], Peng Xiu[6], Jun Hu[2,3], Xin Zhang[4], Feng Zhang[5,8], Haiping Fang[1,2]*

1. School of Science, East China University of Science and Technology, Shanghai 200237, China
2. Shanghai Synchrotron Radiation Facility, Zhangjiang Laboratory (SSRF, ZJLab), Shanghai Advanced Research Institute, Chinese Academy of Sciences, Shanghai 201204, China
3. Shanghai Institute of Applied Physics, Chinese Academy of Sciences, Shanghai 201800, China
4. High Magnetic Field Laboratory, Key Laboratory of High Magnetic Field and Ion Beam Physical Biology, Hefei Institutes of Physical Science, Chinese Academy of Sciences, Hefei 230031, China
5. State Key Laboratory of Respiratory Disease, Guangzhou Institute of Oral Disease, Stomatology Hospital, Department of Biomedical Engineering, School of Basic Medical Sciences, Guangzhou Medical University, Guangzhou 511436, China
6. Department of Engineering Mechanics, Zhejiang University, Hangzhou 310027, China
7. Shanghai Applied Radiation Institute, Shanghai University, Shanghai 200444, China
8. Wenzhou Institute, University of Chinese Academy of Sciences, 16 Xinsan Road, Wenzhou 325001, China
9. University of Chinese Academy of Sciences, Beijing 100049, China

*Corresponding author. Email: fanghaiping@sinap.ac.cn;

**We experimentally demonstrated that the AYFFF self-assemblies adsorbed with various monovalent cations ($Na^+$, $K^+$, and $Li^+$) show unexpectedly super strong paramagnetism. The key to the super strong paramagnetism lies in the hydrated cation–π interactions between the monovalent cations and the aromatic rings in the AYFFF assemblies, which endows the AYFFF-cations complex with magnetic moments.**

There is a large passel of animals navigating using Earth's magnetic field, but it isstill argued that whether the magnetoreception in animals contains enough ferromagnetic components. Recently, we demonstrated that magnesium, zinc, and copper cations on aromatic peptides showed super strong paramagnetism [1]. The key to this unexpected strong paramagnetism lies in the existence of the magnetic moments on the cations adsorbed on aromatic rings in the peptide through hydrated cation-π interactions, where the adsorbed cations display non-divalent behavior with unpaired electron spins and the two dimensional crystal is in the form of CaCl of unconventional stoichiometries [2].

Monovalent cations, such as $Na^+$ in particular, are rich in our body as important electrolytes, which play key roles in osmotic balance, nerve transmission and muscle function [3]. However, monovalent cations form much weaker cation-π interaction with aromatic rings than the divalent cations. Thus, we wonder, for the peptides in which the aromatic rings are not covalently connected with each other, whether the monovalent cations can be stably adsorbed on the individual aromatic rings and thus the aromatic peptide with such cations also show super strong paramagnetism. We noted that on the graphene with polycyclic aromatic rings, there are still two-dimensional $Na_2Cl$ and $Na_3Cl$ crystals of unconventional stoichiometries for monovalent metals, as well as the two-dimensional CaCl and CuCl crystals of unconventional stoichiometries for divalent metals [4].

Here, we show that AYFFF assemblies dispersed in the NaCl, KCl, and LiCl solutions also display super strong paramagnetism. The concentration used here is higher than that of the divalent cations used in Ref. [1].

In our experiment, the AYFFF aromatic peptide powders were first dispersed into pure water with a concentration of 0.5 mg/mL, and stored still at 20 °C for 3 days to form peptide assemblies. The dispersions were then thoroughly mixed with NaCl solutions of different concentrations to obtain mixtures. Then we measured their mass susceptibility ($\chi$) by weighing the magnetic force of peptides at different states under a fixed magnetic field by using a high-precision electronic balance as previously described (1, 5). In brief, we firstly measured the magnetic force per unit mass of the water and peptide assemblies, respectively, and then calculated the $\chi$ of peptides by comparing the value of the magnetic force per unit mass of peptide to that of water, considering that the $\chi$ of pure water is well known as $-7.2 \times 10^{-7}$ emu/g [5].

Figure 1 shows the mass susceptibility ($\chi$) for the peptide in at least three independently repeated trials, together with that of water as a reference. The $\chi$ ranges from $2.45 \times 10^{-5}$ emu/g to $3.93 \times 10^{-4}$ emu/g, with the average value of $1.98 \times 10^{-4}$ emu/g in 150 mM NaCl solution, showing super strong paramagnetism of these peptide assemblies in salt solutions with monovalent cations. Moreover, the average values of the $\chi$ are $2.91 \times 10^{-4}$ emu/g, $5.11 \times 10^{-5}$ emu/g, $5.42 \times 10^{-5}$ emu/g, $7.96 \times 10^{-5}$ emu/g, and $3.95 \times 10^{-5}$ emu/g for the AYFFF peptide assemblies in the KCl and LiCl solutions of 150 mM, and in the NaCl, KCl and LiCl solutions of 40 mM, respectively, suggesting that the super strong paramagnetism of the AYFFF assemblies in the chloride solution of monovalent cations should be universal. It is worth emphasizing that the salt is of critical importance for the observed super strong paramagnetism. In addition, the contribution of salt solutions themselves (at the level of $10^{-7}$ emu/g) to the mass susceptibility of AYFFF assemblies has been subtracted as background during the data analysis. Our further experiments have shown that the AYFFF assemblies in pure water without any salt have a negative average susceptibility of $-3.28 \times 10^{-5}$ emu/g, consistent with the strong diamagnetism observed in our previous work [6].

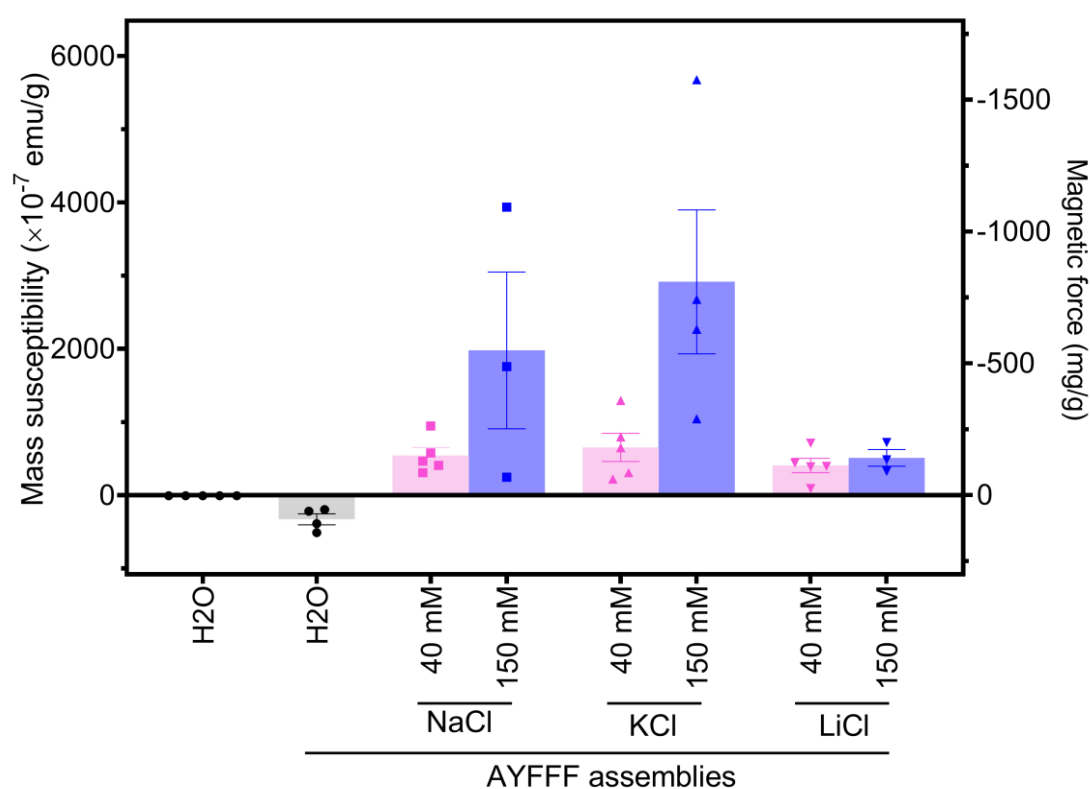

**Figure 1.** *Average mass susceptibility ($\chi$) of AYFFF peptide assemblies in different solutions containing monovalent metal cations, together with water as a reference. The pink circles and blue squares display the experimental data collected in 40 mM and 150 mM salt solutions by the weighing method described in Ref. (5), respectively. Data are shown as mean $\pm$ SEM.*

We attribute the observation of super strong paramagnetism to the monovalent cations adsorbed on aromatic rings through cation-π interactions, similar to the case of divalent cations [1]. We would like to point out that the concentration of monovalent cations used here (150 mM) is in the range of the concentration (~75 to 150 mM) of salt (mostly with monovalent ions) in physiological solutions, which is much higher than that for divalent cations (~40 mM) used in our previous experiment [1]. The cations under such a high concentration have a high probability of interacting with aromatic rings in peptide assemblies, leading to cation-π interactions with the total intensity comparable to those for divalent cations of 40 mM, although cation-π interactions between individual monovalent cation and the aromatic ring is relatively weak. As expected, the mass susceptibilities of AYFFF assemblies is very small in salt solutions with a concentration of 40 mM for the monovalent-cation case (Figure 1).

In summary, the mass susceptibilities obtained by electrical balance experiments clearly demonstrated that the assemblies of an aromatic ring-enriched peptide in the chloride solution of various monovalent cations at the physiologically relevant conditions (room temperature with a salt concentration in the range of the concentration (~75 to 150 mM) of salt in physiological solutions) display super strong paramagnetism. We attribute this phenomenon to the monovalent cations adsorbed on aromatic rings through cation-π interactions. Our findings stimulate novel insights about magnetic effects and magnetic controls on aromatic ring-enriched biomolecules and drugs, including their aggregation, dynamics, delivery and reactions, particularly in living organisms where there are rich monovalent cations, as well as biomaterial fabrication and manipulation.